\documentclass[pre,twocolumn,amsmath,amssymb,floatfix]{revtex4-1}

\usepackage{graphicx}% Include figure files
\usepackage{color}

\begin{document}

\title{Modeling correlated bursts by the bursty-get-burstier mechanism} % Force line breaks with \\
\author{Hang-Hyun Jo}
\email{hang-hyun.jo@apctp.org}
\affiliation{Asia Pacific Center for Theoretical Physics, Pohang 37673, Republic of Korea}
\affiliation{Department of Physics, Pohang University of Science and Technology, Pohang 37673, Republic of Korea}
\affiliation{Department of Computer Science, Aalto University, Espoo FI-00076, Finland}

\date{\today}% It is always \today, today,
             %  but any date may be explicitly specified

\begin{abstract}
    Temporal correlations of time series or event sequences in natural and social phenomena have been characterized by power-law decaying autocorrelation functions with decaying exponent $\gamma$. Such temporal correlations can be understood in terms of power-law distributed interevent times with exponent $\alpha$, and/or correlations between interevent times. The latter, often called correlated bursts, has recently been studied by measuring power-law distributed bursty trains with exponent $\beta$. A scaling relation between $\alpha$ and $\gamma$ has been established for the uncorrelated interevent times, while little is known about the effects of correlated interevent times on temporal correlations. In order to study these effects, we devise the bursty-get-burstier model for correlated bursts, by which one can tune the degree of correlations between interevent times, while keeping the same interevent time distribution. We numerically find that sufficiently strong correlations between interevent times could violate the scaling relation between $\alpha$ and $\gamma$ for the uncorrelated case. A non-trivial dependence of $\gamma$ on $\beta$ is also found for some range of $\alpha$. The implication of our results is discussed in terms of the hierarchical organization of bursty trains at various timescales.
\end{abstract}

%\pacs{89.75.Da,05.40.-a,89.20.-a}
% 89.75.Da Scaling phenomena in complex systems
% 05.40.-a Random processes
% 89.20.-a Interdisciplinary applications of physics

\maketitle

\section{Introduction}\label{sect:intro}

A variety of dynamical processes observed in natural and social phenomena are known to show non-Poissonian or inhomogeneous temporal patterns. Examples include solar flares~\cite{Wheatland1998WaitingTime}, earthquakes~\cite{Corral2004LongTerm, deArcangelis2006Universality}, neuronal firings~\cite{Kemuriyama2010Powerlaw}, and human communication and locomotor activities~\cite{Barabasi2005Origin, Nakamura2007Universal}. Temporal correlations in such time series or event sequences have often been described in terms of $1/f$ noise~\cite{Bak1987Selforganized, Weissman19881f, Ward20071f} or power-law decaying autocorrelation functions~\cite{Karsai2012Universal, Panzarasa2015Emergence}. The autocorrelation function for an event sequence $x(t)$ is defined with delay time $t_d$ as follows:
\begin{equation}
  A(t_d)=\frac{ \langle x(t)x(t+t_d)\rangle_t- \langle x(t)\rangle^2_t}{ \langle x(t)^2\rangle_t- \langle x(t)\rangle^2_t},
\end{equation}
where $\langle\cdot\rangle_t$ means a time average. The event sequence $x(t)$ can be considered to have the value of $1$ at the moment of event occurred, $0$ otherwise. For the event sequences with long-term memory effects, one may find a power-law decaying behavior with decaying exponent $\gamma$:
\begin{equation}
    A(t_d)\sim t_d^{-\gamma}. 
\end{equation}
The autocorrelation function captures the entire temporal correlations present in the event sequence, which can be understood in terms of interevent times and correlations between interevent times. Here the interevent time, denoted by $\tau$, is defined as a time interval between two consecutive events. 

The heterogeneous properties of interevent times have often been characterized by the power-law interevent time distribution $P(\tau)$ with power-law exponent $\alpha$:
\begin{equation}
    \label{eq:Ptau_simple}
    P(\tau)\sim \tau^{-\alpha},
\end{equation}
which may readily imply clustered short interevent times even without correlations between interevent times. This phenomenon has been described in terms of bursts, i.e., rapidly occurring events within short time periods alternating with long inactive periods~\cite{Barabasi2005Origin}. It is well-known that bursty interactions between individuals have a strong influence on the dynamical processes taking place in a network of individuals, such as spreading or diffusion~\cite{Vazquez2007Impact, Karsai2011Small, Miritello2011Dynamical, Rocha2011Simulated, Jo2014Analytically, Delvenne2015Diffusion}. In case when interevent times are fully uncorrelated, i.e., for renewal processes, the power spectral density was analytically calculated from power-law interevent time distributions~\cite{Lowen1993Fractal}. Using this result, one can straightforwardly derive the scaling relation between $\alpha$ and $\gamma$:
\begin{eqnarray}
    \label{eq:alpha_gamma}
    \begin{tabular}{ll}
        $\alpha+\gamma=2$ & for $1<\alpha\leq 2$,\\
        $\alpha-\gamma=2$ & for $2<\alpha\leq 3$.
    \end{tabular}
\end{eqnarray}
This relation was also derived in the context of priority queueing models~\cite{Vajna2013Modelling}. In addition, Abe and Suzuki~\cite{Abe2009Violation} derived $\alpha+p=2$ for $1<\alpha<2$ and $0<p<1$, where the scaling exponent $p$ characterizes Omori law in earthquakes, with a possible interpretation of $p=\gamma$.

Then a natural question arises: If the interevent times are correlated with each other, would the above scaling relation still hold? In other words, in order to violate the above scaling relation, how strong correlations between interevent times should be introduced? In order to study this question, the correlations between interevent times can be measured using the notion of bursty trains~\cite{Karsai2012Universal}. A bursty train is defined as a set of events such that interevent times between any two consecutive events in the bursty train are less than or equal to a given time window $\Delta t$, while those between events in different bursty trains are larger than $\Delta t$. The number of events in the bursty train is called burst size, and it is denoted by $b$. The distribution of $b$ follows an exponential function if the interevent times are fully uncorrelated with each other. However, $b$ has been empirically found to be power-law distributed, i.e.,
\begin{equation}
    \label{eq:burstSizeDistribution}
    P_{\Delta t}(b)\sim b^{-\beta} 
\end{equation}
for a wide range of $\Delta t$, e.g., in earthquakes, neuronal activities, and human communication patterns~\cite{Karsai2012Universal, Karsai2012Correlated, Wang2015Temporal}. This indicates the presence of correlations between interevent times, hence this phenomenon is called \emph{correlated bursts}. We also note that the exponential distributions of $P_{\Delta t}(b)$ have recently been reported for mobile phone calls of individual users in another work~\cite{Jiang2016Twostate}. In sum, we expect that the temporal correlations observed in the autocorrelation function $A(t_d)$ can be fully understood in terms of the statistical properties of interevent times, e.g., $P(\tau)$, together with those of the correlations between interevent times, e.g., $P_{\Delta t}(b)$. Simply put, one can study the dependence of $\gamma$ on $\alpha$ and $\beta$, which is the aim of this paper.

The dependence of $\gamma$ on $\alpha$ and $\beta$ has been investigated by dynamically generating event sequences showing temporal correlations, described by the power-law distributions of interevent times and burst sizes in Eqs.~(\ref{eq:Ptau_simple}) and~(\ref{eq:burstSizeDistribution}). These generative approaches were based on either two-state Markov chain~\cite{Karsai2012Universal} or self-exciting point processes~\cite{Jo2015Correlated}. Although such approaches have been successful for reproducing empirically-observed correlated bursts to some extent, our understanding of the underlying mechanisms for correlated bursts is far from complete. In addition to the generative modeling approaches, here we take an alternative approach by devising the bursty-get-burstier model, where power-law distributions of interevent times and burst sizes are inputs rather than outputs of the model. In our model, one can explicitly tune the degree of correlations between interevent times to test whether the scaling relation in Eq.~(\ref{eq:alpha_gamma}) will be violated due to the correlations between interevent times.

Our paper is organized as follows: In Sect.~\ref{sect:model}, the bursty-get-burstier model is devised to simulate event sequences with tunable correlations between interevent times, and to study the effects of such correlations on the scaling relation established for the uncorrelated interevent times. We conclude our work in Sect.~\ref{sect:concl}.

\section{Methods and results}\label{sect:model}

\subsection{Uncorrelated interevent times}\label{subsect:uncorrelated}

As a baseline, we investigate the case with uncorrelated interevent times. We first relate power-law exponents characterizing bursty time signals as mentioned in Sect.~\ref{sect:intro}. The power spectral density $P(f)$ of a time signal $x(t)$, where $f$ denotes the frequency, is known to be the Fourier transform of the autocorrelation function $A(t_d)$. Thus if $A(t_d) \sim t_d^{-\gamma}$ for $0<\gamma<1$, then one finds the scaling of $P(f)\sim f^{-\eta}$ with 
\begin{equation}
    \label{eq:eta_gamma}
    \eta=1-\gamma. 
\end{equation}
When the interevent times are i.i.d. random variables with $P(\tau)\sim \tau^{-\alpha}$, implying no correlations between interevent times, the power-law exponent $\eta$ is obtained as a function of $\alpha$ as follows~\cite{Lowen1993Fractal, Allegrini2009Spontaneous}:
\begin{equation}
    \label{eq:eta_alpha}
    \eta=\left\{\begin{tabular}{ll}
            $\alpha-1$ & for $1<\alpha\leq 2$,\\
            $3-\alpha$ & for $2<\alpha\leq 3$.
        \end{tabular}\right.
\end{equation}
Combining Eqs.~(\ref{eq:eta_gamma}) and~(\ref{eq:eta_alpha}), we obtain Eq.~(\ref{eq:alpha_gamma}): $\gamma$ decreases with $\alpha$ for $1<\alpha\leq 2$ and then it increases with $\alpha$ for $2<\alpha\leq 3$. These power-law exponents can also be related via Hurst exponent $H$, i.e., $\gamma=2-2H$~\cite{Kantelhardt2001Detecting, Rybski2009Scaling} or $\eta=2H-1$~\cite{Allegrini2009Spontaneous, Rybski2012Communication}.

In order to numerically study the case with uncorrelated interevent times, an event sequence is constructed from a prescribed interevent time distribution, as done in Ref.~\cite{Lowen1993Fractal}. A similar approach for bursty dynamics has been studied~\cite{Perotti2014Temporal, Kim2016Measuring}, where events are randomly distributed in time, e.g., to simulate the regular, random, and extremely bursty time series~\cite{Kim2016Measuring}. Here we construct an event sequence $x(t)$ with $n+1$ events using $n$ uncorrelated interevent times, $\{\tau_1,\cdots,\tau_n\}$, that are independently drawn from $P(\tau)$. Let us assume that the zeroth event occurs at $t_0=0$. Then the timing of the $i$th event for $i=1,\cdots,n$ is given by $t_i=\sum_{i'=1}^i \tau_{i'}$, leading to the event sequence $x(t)$ in the range of $0\leq t \leq t_n$ as
\begin{equation}
    \label{eq:x_t}
    x(t)=\left\{\begin{tabular}{ll}
            $1$ & for $t\in\{t_i\}$,\\
            $0$ & otherwise.
        \end{tabular}\right.
\end{equation}

\begin{figure}[!t]
    \includegraphics[width=\columnwidth]{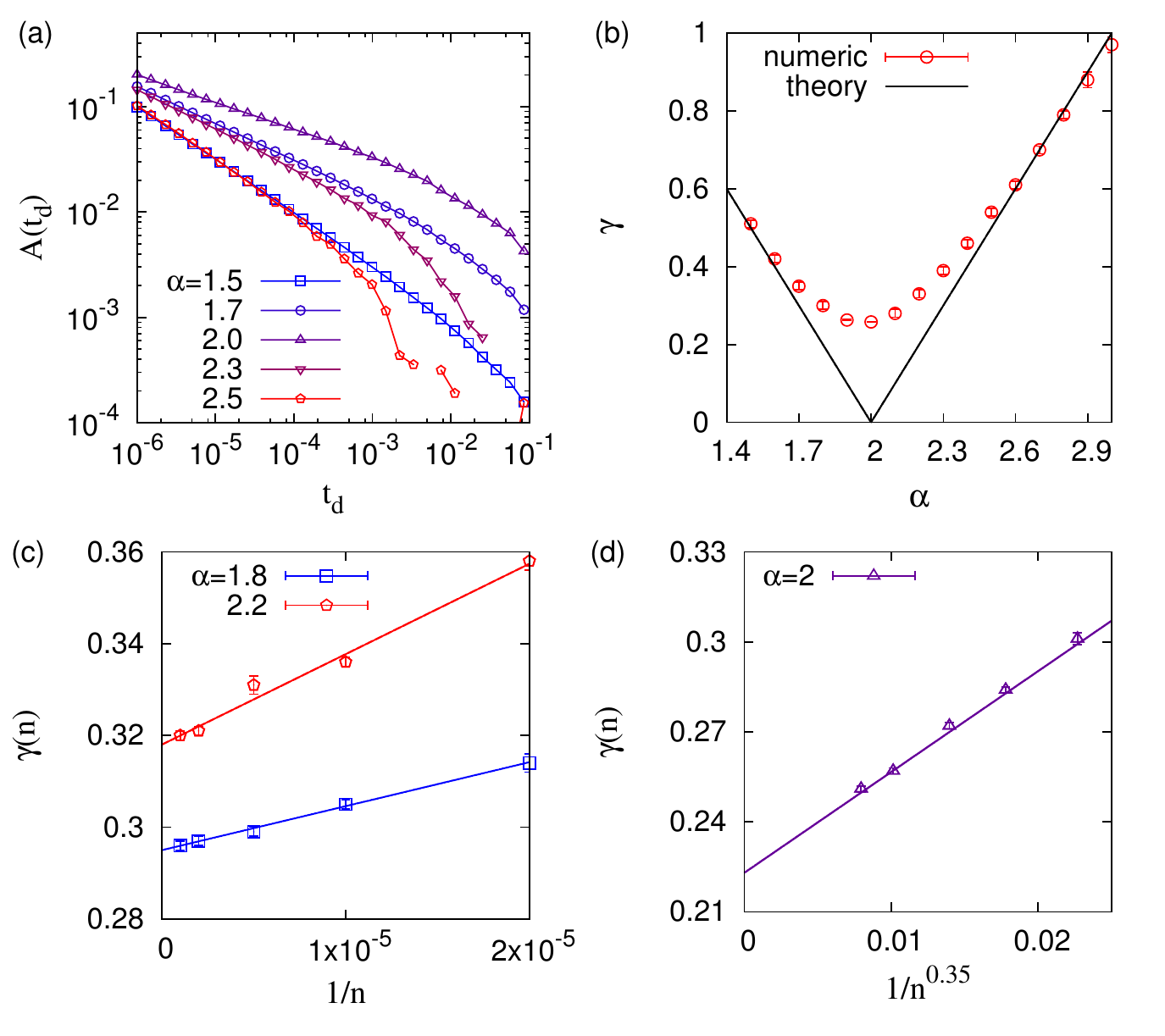}
    \caption{(Color online) Numerical test of the scaling relation in Eq.~(\ref{eq:alpha_gamma}) for the event sequences constructed using uncorrelated interevent times, that are drawn from $P(\tau)$ in Eq.~(\ref{eq:Ptau}). (a) Autocorrelation functions $A(t_d)$ for various values of $\alpha$. For each value of $\alpha$, we have generated $50$ event sequences with $n=5\cdot 10^5$ and $\tau_0=10^{-7}$. (b) The estimated value of $\gamma$ as a function of $\alpha$ using $A(t_d)\sim t_d^{-\gamma}$ (red circles), compared to the theory in Eq.~(\ref{eq:alpha_gamma}) (solid line). (c) Finite-size effects on the values of $\gamma(n)$ for $\alpha=1.8$ and $2.2$, fitted with the form of $1/n$. (d) The values of $\gamma(n)$ for $\alpha=2$ are better described by the form of $1/n^{0.35}$ than by that of $1/n$.}
    \label{fig:uncorrelated}
\end{figure}

We adopt the power-law interevent time distribution with $\alpha>1$ as follows:
\begin{equation}
    \label{eq:Ptau}
    P(\tau)= \left\{\begin{tabular}{ll}
    $(\alpha-1)\tau_0^{\alpha-1}\tau^{-\alpha}$ & for $\tau\geq \tau_0$,\\
            $0$ & otherwise,
        \end{tabular}\right.\\
\end{equation}
where $\tau_0$ is the lower bound of interevent times. For each value of $\alpha$, $50$ event sequences are generated with $n=5\cdot 10^5$ and $\tau_0=10^{-7}$~\footnote{In order to avoid too large interevent times, especially for small $\alpha$, we introduce the upper bound of interevent times for numerical simulations. This upper bound is set to be $1$, hence being believed to have a negligible effect on the results.}. As shown in Fig.~\ref{fig:uncorrelated}, we find that the numerical value of $\gamma$ as a function of $\alpha$ is comparable to the scaling relation in Eq.~(\ref{eq:alpha_gamma}) for $\alpha\lesssim 1.6$ and $\alpha\gtrsim 2.4$. The discrepancy between the theoretical and numerical values for $1.6\lesssim \alpha\lesssim 2.4$ could be due to the logarithmic correction appearing around at $\alpha=2$ as well as due to finite-size effects. The finite-size effects can be studied by measuring the $n$-dependence of $\gamma$ as shown in Fig.~\ref{fig:uncorrelated}(c), where $\gamma(n)=\gamma(\infty)+ cn^{-1}$ with constant $c$ is used to estimate $\gamma(\infty)=0.295(1)$ for $\alpha=1.8$ and $\gamma(\infty)=0.318(2)$ for $\alpha=2.2$, respectively. Both limiting values of $\gamma(\infty)$ are yet different from $0.2$ that is expected from the scaling relation in Eq.~(\ref{eq:alpha_gamma}). In case with $\alpha=2$ [see Fig.~\ref{fig:uncorrelated}(d)], we use the form of $\gamma(n)=\gamma(\infty)+cn^{-0.35}$ to estimate $\gamma(\infty)=0.223(8)$. These deviations from the theoretical values may imply that other effects like logarithmic corrections could be strong. Instead of studying such effects in more detail, we will consider the fixed number of $n=5\cdot 10^5$ for the rest of our paper because this number is already large enough to study the temporal properties of most empirical datasets. In addition, for all cases, we obtain the exponential burst size distribution $P_{\Delta t}(b)\propto e^{-b/b_c(\Delta t)}$ with exponential cutoff $b_c(\Delta t)$ for the wide range of $\Delta t$ (not shown), as expected.

The decreasing and then increasing behavior of $\gamma$ as a function of $\alpha$ can be roughly understood by the fact that the autocorrelation function essentially measures the chance of finding two events occurred in $t$ and $t+t_d$, no matter how many events occur between those two events. This chance can be written as~\cite{Lowen1993Fractal}
\begin{equation}
    \label{eq:auto_series}
    \langle x(t)x(t+t_d)\rangle_t \propto \sum_{k=1}^\infty P^{\star k}(t_d)\equiv G(t_d),
\end{equation}
where $P^{\star k}(t_d)$ denotes the probability of finding $k$ consecutive interevent times whose sum is exactly equal to $t_d$. The term with $k=1$ in Eq.~(\ref{eq:auto_series}) simply corresponds to the value of the interevent time distribution at $\tau=t_d$. This value is proportional to $(\alpha-1)(\frac{\tau_0}{t_d})^{\alpha}$ that is increasing and then decreasing for $t_d>\tau_0$ as $\alpha$ varies from $1$ to $3$. Such non-monotonic behavior is expected to other terms with $k>1$ as well, see Appendix~\ref{append:convol} for the details. Then combining all terms, $G(t_d)$ can have the maximum value for an intermediate range of $\alpha$. With an additional assumption that the larger values of $G(t_d)$ for a wide range of $t_d$ imply the smaller $\gamma$, one can get some hints on why $\gamma$ has the lowest value around at $\alpha=2$. 

\subsection{Correlated interevent times}\label{subsect:correlated}

In order to implement the correlations between interevent times, we devise the bursty-get-burstier model of correlated interevent times, where the power-law distributions of interevent times and burst sizes are inputs rather than outputs of the model. Along with the power-law form of $P(\tau)$ in Eq.~(\ref{eq:Ptau}), we adopt power-law burst size distributions as $P_{\Delta t}(b)\sim b^{-\beta}$ for a wide range of $\Delta t$, meaning that an event sequence will be constructed from the prescribed power-law distributions of interevent times and burst sizes. As in the uncorrelated case, we prepare the set of $n$ interevent times, $T\equiv \{\tau_1,\cdots,\tau_n\}$, that are independently drawn from $P(\tau)$. Correlations between interevent times are implemented by shuffling or permuting those interevent times according to their sizes, implying that only the ordering of interevent times in $T$ is affected. Each permutation will result in each realized event sequence but from exactly the same $T$.

\begin{figure}[!t]
    \includegraphics[width=\columnwidth]{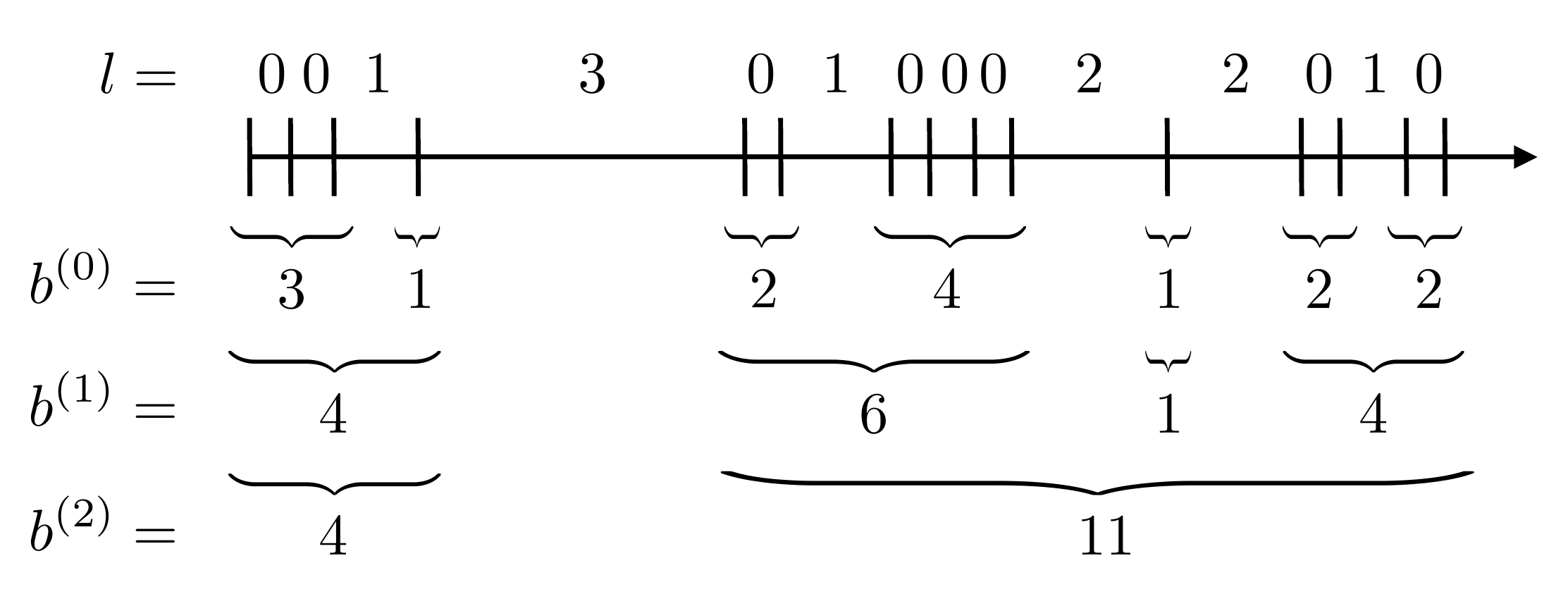}
    \caption{Schematic diagram for the hierarchical organization of burst trains at different timescales or ``levels'', with $15$ events, denoted by vertical lines, and $14$ interevent times. The numbers above the event sequence indicate the levels of interevent times, while $b^{(l)}$s are the burst sizes identified using $\Delta t_l$. For the definition of $\Delta t_l$, see the main text.}
    \label{fig:model}
\end{figure}

\begin{figure}[!t]
    \includegraphics[width=\columnwidth]{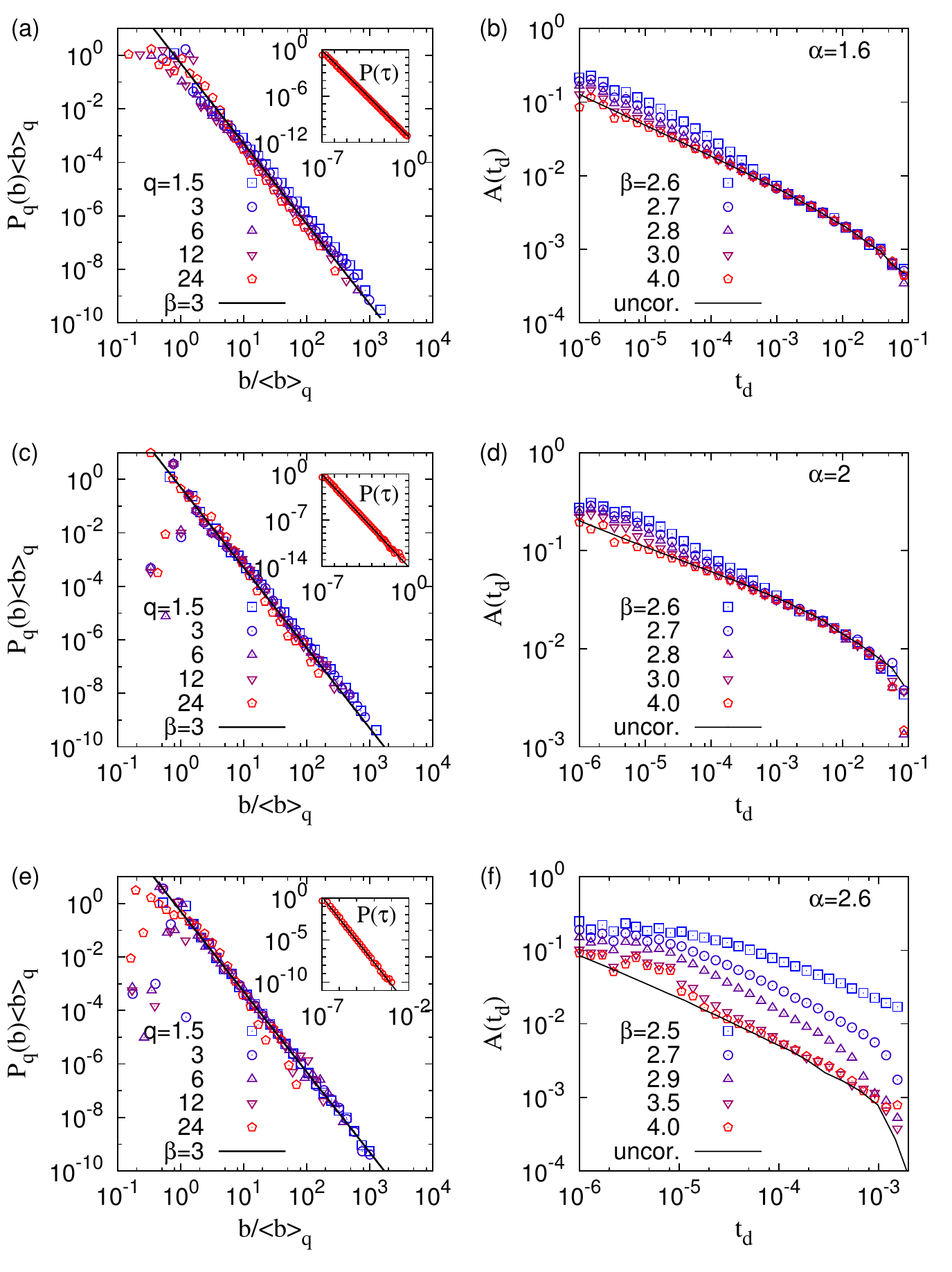}
    \caption{(Color online) Numerical results of the bursty-get-burstier (BGB) model of correlated interevent times for various values of $\alpha$ and $\beta$. We show the cases with $\alpha=1.6$ (top), $\alpha=2$ (middle), and $2.6$ (bottom). In (a), (c), and (e), we numerically confirm our BGB method for making burst size distributions with $\beta=3$ for various time windows of $\Delta t= \tau_0 q$ with $q=q_0s^l$. $\langle b\rangle_q$ is the average burst size of bursty trains identified using $\Delta t=\tau_0 q$. Insets show the corresponding interevent time distributions. (b), (d), and (f) show the autocorrelation functions for different values of $\beta$ (colored points), compared to the corresponding uncorrelated cases (black curves). For each curve, we have generated up to $100$ event sequences with $n=5\cdot 10^5$, $\tau_0=10^{-7}$, $q_0=1.5$, $s=2$, and $L=4$.}
    \label{fig:correlated_auto}
\end{figure}

In order to impose power-law distributed burst sizes for a wide range of $\Delta t$, we first partition $T$ into several subsets, denoted by $T_l$, at different timescales or ``levels'' $l=0,1,\cdots,L$:
\begin{eqnarray}
    T_0 &\equiv & \{\tau_i| \tau_0 \leq \tau_i \leq \Delta t_0\}, \nonumber\\
    T_l &\equiv & \{\tau_i| \Delta t_{l -1} < \tau_i \leq \Delta t_l \}\ \mbox{for}\ l=1,\cdots,L-1, \\
    T_L &\equiv & \{\tau_i| \tau_i > \Delta t_{L-1} \},\nonumber
\end{eqnarray}
where $\Delta t_l\equiv \tau_0 q_0s^l$ with constants $q_0,s > 1$ is the size of the time window at the level $l$. This partition with some constants $q_0$ and $s$ readily determines the number of bursty trains, denoted by $m_l$, that would be identified if $\Delta t_l$ were used as the time window, no matter which permutation for $T$ is chosen for constructing the event sequence. It is because each burst size, say $b^{(l)}$, implies $b^{(l)}-1$ consecutive interevent times less than or equal to $\Delta t_l$ and one interevent time larger than $\Delta t_l$. Precisely, we get
\begin{equation}
    m_l = \big|\cup_{l'=l+1}^{L} T_{l'}\big| + 1.
\end{equation}
Let us now denote the sizes of bursty trains using $\Delta t_l$ by $B_l\equiv \{b^{(l)}\}$, with $m_l=|B_l|$. Here the burst sizes, $b^{(l)}$s, are to follow a power law with the same exponent $\beta$ at all levels, for which bursty trains must be hierarchically organized as schematically depicted in Fig.~\ref{fig:model}. For this, we adopt the power-law burst size distribution only at the level $l=0$:
\begin{equation}
    \label{eq:Pb}
    P_{\Delta t_0}(b^{(0)}) = \frac{1}{\zeta(\beta)} b^{(0)-\beta}\ \textrm{for}\ b^{(0)}=1,2,\cdots,
\end{equation}
where $\zeta(\cdot)$ denotes the Riemann zeta function. Then $m_0$ burst sizes are independently drawn from $P_{\Delta t_0}(b^{(0)})$ in Eq.~(\ref{eq:Pb}) to obtain $B_0$. Note that the sum of burst sizes in $B_0$ must be $n+1$. As depicted in Fig.~\ref{fig:model}, for a given $B_l$ with $l\geq 0$, $B_{l+1}$ can be constructed by merging several $b^{(l)}$s to make each $b^{(l+1)}$, but under the condition that those $b^{(l+1)}$s are to be power-law distributed with the same exponent $\beta$ in Eq.~(\ref{eq:Pb}). This condition can be satisfied if bigger bursts tend to be merged with bigger ones, and smaller bursts with smaller ones. In other words, bigger (smaller) bursts are followed by bigger (smaller) ones, hence this merging rule can be called the bursty-get-burstier (BGB) method. Precisely, we devise the following method: $B_l$ is sorted, e.g., in a descending order, then it is sequentially partitioned into $m_{l+1}$ subsets of the (almost) same size. The size of each subset may be either $\lfloor \frac{m_{l}}{m_{l+1}}\rfloor$ or $\lfloor \frac{m_{l}}{m_{l+1}}\rfloor+1$. The sum of $b^{(l)}$s in each subset leads to one $b^{(l+1)}$. This procedure is repeated until $l=L-1$. We numerically confirm that our BGB method indeed generates power-law tails in distributions of $b^{(l)}$s with the same exponent $\beta$ at all levels, as shown in Fig.~\ref{fig:correlated_auto}(a,c,e) for the case with $\beta=3$.

\begin{figure}[!t]
    \includegraphics[width=\columnwidth]{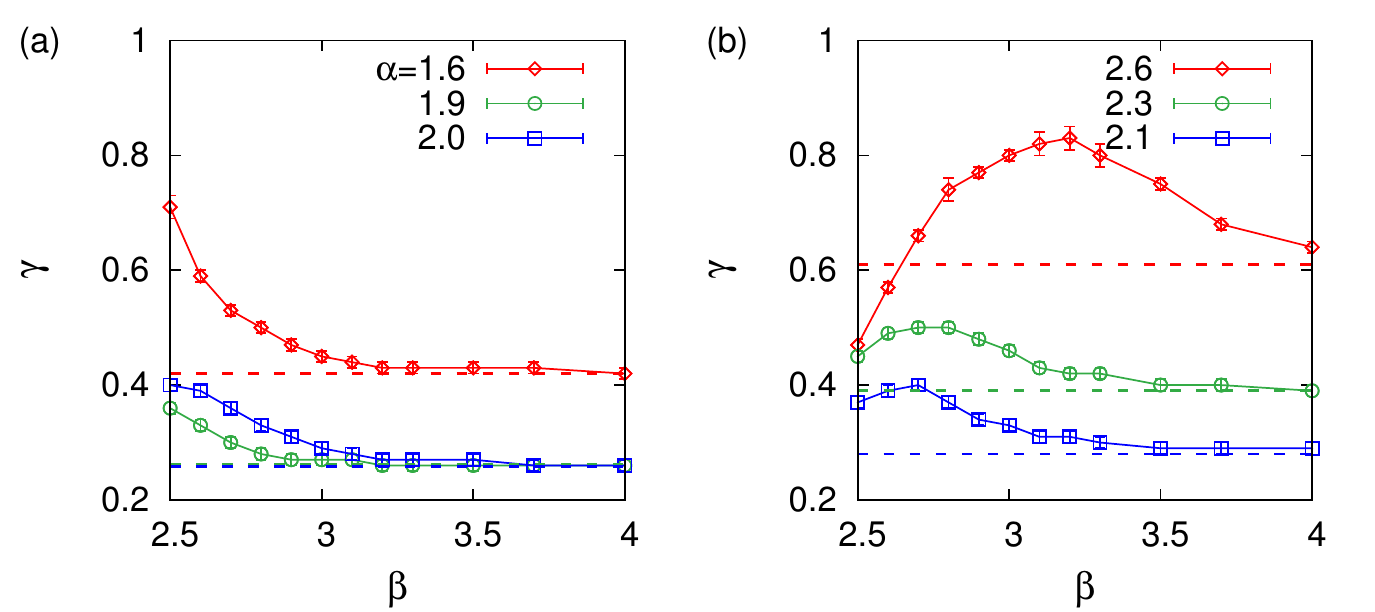}
    \caption{(Color online) The values of $\gamma$ measured from autocorrelation functions, e.g., in Fig.~\ref{fig:correlated_auto}, for various values of $\alpha$ and $\beta$, with horizontal dashed lines corresponding to those for the uncorrelated cases measured in Fig.~\ref{fig:uncorrelated}(b).}
    \label{fig:correlated_gamma}
\end{figure}

We then show how to construct the event sequence $x(t)$ by permuting interevent times in $T$, based on information about which bursty trains at the level $l$ have been merged into which bursty train at the level $l+1$ for all possible $l$s. We begin with the highest level $L-1$ by constructing a sequence of burst sizes at the level $L-1$ in a random order, alternating with interevent times randomly drawn from $T_L$ without replacement, denoted by $\tau^{(L)}$:
\begin{equation}
    (b^{(L-1)}_1, \tau^{(L)}_1, b^{(L-1)}_2, \tau^{(L)}_2, \cdots, b^{(L-1)}_{m_{L-1}}). 
\end{equation}
Note that from now on, all the subscripts are dummy indexes. Then, each $b^{(L-1)}$ is replaced by a sequence made of $b^{(L-2)}$s, which have been merged together to make the $b^{(L-1)}$, in a random order, alternating with interevent times randomly drawn from $T_{L-1}$ without replacement. This procedure is repeated for all bursty trains at all levels, eventually resulting in a sequence made of exactly $n$ interevent times. From this sequence of interevent times, the event timings are given by $t_0=0$ and $t_i=\sum_{i'=1}^i \tau_{i'}$ for $i=1,\cdots,n$, then we finally obtain the event sequence $x(t)$ by Eq.~(\ref{eq:x_t}).

Once $x(t)$ is generated, we first measure distributions of interevent times and burst sizes for various values of $\Delta t$ or $q=\frac{\Delta t}{\tau_0}$ to confirm our BGB method, and then we calculate autocorrelation functions to test whether the scaling relation in Eq.~(\ref{eq:alpha_gamma}) holds or not in the presence of the correlations between interevent times. In Fig.~\ref{fig:correlated_auto}, we show the numerical results for various values of $\alpha$ and $\beta$, where we have used $\tau_0=10^{-7}$, $q_0=1.5$, $s=2$, and $L=4$. We find that $P(\tau)\sim \tau^{-\alpha}$ and $P_{\Delta t}(b)\sim b^{-\beta}$ for a wide range of $\Delta t$ as expected, see Fig.~\ref{fig:Pb} for more details. We remark that regarding our setting for the power-law burst size distribution at $l=0$ in Eq.~(\ref{eq:Pb}), one can show that the value of $\beta$ is determined for given $\alpha$, $q_0$, and $n$, as analyzed in Appendix~\ref{append:beta}. However, by assuming that it is sufficient to show power-law tails in the burst size distributions, we can simulate a wide range of $\beta$ using our BGB method.

Then, for each $\alpha$, autocorrelation functions $A(t_d)$ for various values of $\beta$ are compared to that for the uncorrelated case, e.g., in Fig.~\ref{fig:correlated_auto}(b,d,f). The estimated values of $\gamma$ for various values of $\alpha$ and $\beta$ are presented in Fig.~\ref{fig:correlated_gamma}. When $\alpha\leq 2$, it is numerically found that the autocorrelation functions for $\beta\lesssim 3$ deviate from the uncorrelated case, implying the violation of scaling relation between $\alpha$ and $\gamma$ in Eq.~(\ref{eq:alpha_gamma}). Precisely, the smaller $\beta$ leads to the larger $\gamma$, implying that the stronger correlations between interevent times may induce the faster decaying of autocorrelation. The deviation observed for $\beta\lesssim 3$ could be due to the fact that the variance of $b$ diverges for $\beta<3$. On the other hand, in case with $\alpha > 2$, the estimated $\gamma$ deviates significantly from that for the uncorrelated case for the almost entire range of $\beta$, although $\gamma$ seems to approach the uncorrelated case as $\beta$ increases. Interestingly, the estimated values of $\gamma$ show an increasing and then decreasing behavior as $\beta$ increases. The reason for such different behaviors of $\gamma$ for $\alpha\leq 2$ and for $\alpha > 2$ can be rooted in the non-monotonic behavior of $\gamma$ as a function of $\alpha$ in the uncorrelated cases. In order to understand this difference, more rigorous studies are needed in a future.

\section{Conclusion}\label{sect:concl}

The effects of correlations between interevent times, often called correlated bursts, on the temporal correlations characterized by power-law decaying autocorrelation functions, are far from being fully understood. In order to study these effects systematically, we have devised the bursty-get-burstier (BGB) model, where power-law distributions of interevent times and burst sizes are inputs rather than outputs of the model. With our model, one can tune the degree of correlations between interevent times, while keeping the same interevent time statistics. Then the established scaling relation between power-law exponent $\alpha$ for interevent time distributions and decaying exponent $\gamma$ for autocorrelation functions can be numerically tested, especially, whether the scaling relation can be violated due to the correlations between interevent times. 

As a baseline, we numerically study the case of uncorrelated interevent times. We find that the dependence of $\gamma$ on $\alpha$ is comparable to the theoretical expectation in Eq.~(\ref{eq:alpha_gamma}) for the range of $\alpha$ far from $2$. It is because for $\alpha\approx 2$, the logarithmic corrections become effective. Next, using our BGB model, we generate the event sequences showing power-law distributions of both interevent times and burst sizes for a wide range of the time window. By measuring the autocorrelation functions of those event sequences and then by estimating the decaying exponents of them, we find that the correlations between interevent times can violate the scaling relation established for the uncorrelated case, but in different ways depending on the range of $\alpha$.

Our BGB model for correlated bursts turns out to be useful for imposing power-law distributed burst sizes at different timescales simultaneously, which is however not trivial to implement. It is because it requires somewhat deliberate hierarchical organization of burst sizes at different timescales, namely the bursty-get-burstier mechanism. There can be other ways of implementing such hierarchical organization. We can get some hints from this hierarchical organization for the origin of correlated bursts. For example, if human communication patterns can be described in terms of correlated bursts, human individuals might have organized their communication activities in a hierarchical way either consciously or unconsciously: Bigger bursts tend to be followed by bigger ones, while smaller bursts follow smaller ones, at all relevant timescales of human dynamics.

\begin{acknowledgments}
    The author thanks Mikko Kivel\"a, J\'anos Kert\'esz, and Kimmo Kaski for fruitful discussions, and he acknowledges financial support by Basic Science Research Program through the National Research Foundation of Korea (NRF) grant funded by the Ministry of Education (2015R1D1A1A01058958).
\end{acknowledgments}

\appendix

\begin{figure*}[!t]
    \includegraphics[width=2\columnwidth]{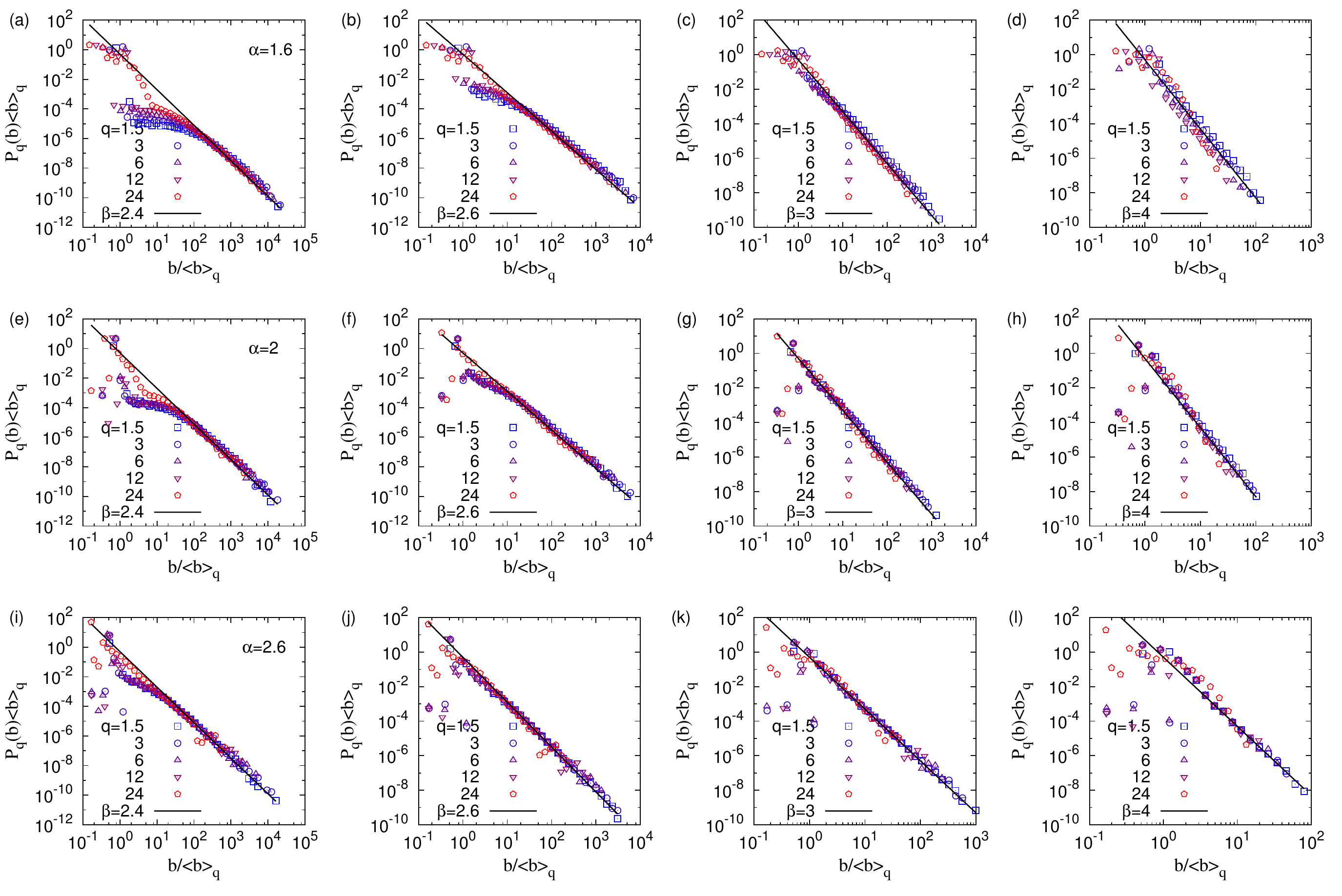}
    \caption{(Color online) Burst size distributions of the bursty-get-burstier model of correlated interevent times for $\alpha=1.6$ (top), $2$ (middle), and $2.6$ (bottom), for $\beta=2.4$, $2.6$, $3$, and $4$ (from left to right), and for various values of $\Delta t= \tau_0 q$ with $q=q_0s^l$. $\langle b\rangle_q$ denotes the average burst size of bursty trains identified using $\Delta t=\tau_0 q$. For each case, we have generated $20$ event sequences with $n=5\cdot 10^5$, $\tau_0=10^{-7}$, $q_0=1.5$, $s=2$, and $L=4$.}
    \label{fig:Pb}
\end{figure*}

\section{Asymptotic result of the convolution of the interevent time distribution}\label{append:convol}

In Eq.~(\ref{eq:auto_series}), $P^{\star k}(t_d)$ denotes the $k$th order convolution of the interevent time distribution $P(\tau)$, i.e., $P^{\star 1}(t_d)=P(\tau=t_d)$ and $P^{\star k}(t_d)=\int_0^{t_d}P(\tau)P^{\star k-1}(t_d-\tau)d\tau$ for $k>1$. In other words, one can write as follows~\cite{Jo2013Contextual}:
\begin{equation}
    P^{\star k}(t_d)=\prod_{i=1}^k\int_0^\infty d\tau_i P(\tau_i)\cdot \delta\left(t_d-\sum_{i=1}^k \tau_i\right).
\end{equation}
Here we analyze $P^{\star k}(t_d)$ for $t_d\gg \tau_0$. By taking a Laplace transform of $P^{\star k}(t_d)$ with respect to $t_d$, one gets
\begin{equation}
    \widetilde{P^{\star k}}(s) = \tilde{P}(s)^k,
\end{equation}
where $\tilde P(s)$ denotes the Laplace transform of $P(\tau)$ in Eq.~(\ref{eq:Ptau}) and is given by
\begin{equation}
    \tilde{P}(s)=(\alpha-1)(\tau_0 s)^{\alpha-1} \Gamma(1-\alpha,\tau_0s).
\end{equation}
The incomplete Gamma function is expanded in the asymptotic limit of $s\to 0$ to obtain
\begin{equation}
    \label{eq:convol_series}
    \tilde{P}(s)\approx 1 + \Gamma(1-\alpha) (\alpha-1)(\tau_0 s)^{\alpha-1} -\frac{\alpha-1}{\alpha-2}\tau_0s +\cdots.
\end{equation}
If $1<\alpha<2$, since the term of the order of $s^{\alpha-1}$ dominates that of $s$, we can ignore the higher order terms to get
\begin{eqnarray}
    \tilde{P}(s)^k &\approx& 1 + k \Gamma(1-\alpha) (\alpha-1)(\tau_0 s)^{\alpha-1}\\
     &=& 1 + \Gamma(1-\alpha) (\alpha-1)(\tau_k s)^{\alpha-1},
\end{eqnarray}
where we have defined $\tau_k$ by $\tau_k^{\alpha-1} \equiv k\tau_0^{\alpha-1}$. That is, the $k$th convolution can interpreted as the replacement of $\tau_0$ by $\tau_k$. We finally get for $t_d\gg \tau_0$
\begin{eqnarray}
    P^{\star k}(t_d) &\approx& (\alpha-1)\tau_k^{\alpha-1} t_d^{-\alpha}\\
    &=& k (\alpha-1)\tau_0^{\alpha-1} t_d^{-\alpha},
\end{eqnarray}
which turns out to increase as $\alpha$ varies from $1$. If $\alpha>2$, since the term of the order of $s$ in Eq.~(\ref{eq:convol_series}) becomes dominant, we can similarly obtain
\begin{eqnarray}
    \tilde{P}(s)^k &\approx& 1 -k \frac{\alpha-1}{\alpha-2}\tau_0s,\\
    P^{\star k}(t_d) &\approx& k^{\alpha-1}(\alpha-1)\tau_0^{\alpha-1} t_d^{-\alpha}.
\end{eqnarray}
This $P^{\star k}(t_d)$ must be decreasing according to $\alpha$ as long as $\frac{k\tau_0}{t_d}<1$. In sum, the convolution of the interevent time distribution is increasing for small $\alpha$ and decreasing for large $\alpha$, implying the non-monotonic behavior of $P^{\star k}(t_d)$ according to $\alpha$.

\section{Exact scaling relation between $\alpha$ and $\beta$}\label{append:beta}

Here we show that in principle, $\alpha$ and $\beta$ cannot be independent of each other. As mentioned in the main text, the sum of burst sizes in $B_0$ must be equal to the total number of events, i.e., $n+1$, implying that
\begin{equation}
    \label{eq:sum_burst}
    m_0\langle b^{(0)}\rangle=n+1,
\end{equation}
where $\langle b^{(0)}\rangle$ denotes the average burst size. Note that this equation holds for the arbitrary choice of $P(\tau)$ and $P_{\Delta t_0}(b^{(0)})$. In our setting with Eqs.~(\ref{eq:Ptau}) and (\ref{eq:Pb}), since
\begin{eqnarray}
    m_0 &=& n\int_{\Delta t_0}^\infty P(\tau)d\tau = n q_0^{1-\alpha},\\
    \langle b^{(0)}\rangle &=& \sum_{b^{(0)}=1}^\infty b^{(0)}P_{\Delta t_0}(b^{(0)}) = \frac{\zeta(\beta-1)}{\zeta(\beta)},
\end{eqnarray}
we obtain the relation between $\alpha$ and $\beta$ from Eq.~(\ref{eq:sum_burst}) as follows:
\begin{equation}
    \frac{\zeta(\beta-1)}{\zeta(\beta)} = \frac{n+1}{n} q_0^{\alpha-1}.
\end{equation}
This result can be seen as another scaling relation between $\alpha$ and $\beta$ for given $q_0$ and $n$. For example, when $q_0=1.5$ and $n\gg 1$, one gets $\beta\approx 2.81$ for $\alpha=2$, and $\beta\approx 2.51$ for $\alpha=2.6$, respectively. We remark that this relation is based on the assumption that the burst size distribution follows a clear power law for the entire range of $b^{(0)}$. On the other hand, in practice, we can simulate a much wider range of $\beta$ by considering the distributions showing power laws only in their tails.

%\bibliography{/Users/h2jo/Research/_papers/h2jo-papers}
\bibliographystyle{apsrev4-1}

\end{document}